\documentclass[12pt]{article}\usepackage[hyperfootnotes=false]{hyperref}
\usepackage{epsfig}
\usepackage{float}
\usepackage{empheq}
\usepackage{bbold}

\usepackage{xspace}

\makeatletter

\@addtoreset{equation}{section}
\makeatother
\usepackage[utf8]{inputenc}
\usepackage{amsmath}
\usepackage{caption}
\usepackage{amsmath}
\usepackage{amssymb}
\usepackage{graphicx}

\newlength{\abstractwidth}
\renewcommand{\thefootnote}{\fnsymbol{footnote}}
\renewcommand{\thanks}[1]{\footnote{#1}} % Use this for footnotes
\newcommand{\starttext}{
\setcounter{footnote}{0}
\renewcommand{\thefootnote}{\arabic{footnote}}}

\newcommand{\be}{\begin{equation}}
\newcommand{\bea}{\begin{eqnarray}}
\newcommand{\eea}{\end{eqnarray}}
\newcommand{\beq}{\begin{equation}}
\newcommand{\ee}{\end{equation}}

	\newcommand*\widefbox[1]{\fbox{\hspace{2em}#1\hspace{2em}}}

	\def\dsp.{de Sitter space.}
	\def\eq{&=&}

	\def\ra{\rangle}
	\def\simleq{\; \raise0.3ex\hbox{$<$\kern-0.75em
			\raise-1.1ex\hbox{$\sim$}}\; }
	\def\simgeq{\; \raise0.3ex\hbox{$>$\kern-0.75em
			\raise-1.1ex\hbox{$\sim$}}\; }
	
	\def\bi{\begin{itemize}}
		\def\ei{\end{itemize}}

	\def\dof{degrees of freedom }

	\def\CO{{\cal{O}}}

	\def\CT{{\cal{T}}}

	\def\Tr{\rm Tr \it}
	\def\bsub{ \begin{subequations}
			\begin{empheq}[box=\widefbox]{align}  }
			\def\esub{ \end{empheq}
	\end{subequations}}
	
	\def\1{\(  \mathbb{1} \)}

	\def\bn{\bigskip \noindent}

	\makeatletter
	\g@addto@macro\normalsize{%
		\setlength\abovedisplayskip{10pt}
		\setlength\belowdisplayskip{20pt}
		\setlength\abovedisplayshortskip{10pt}
		\setlength\belowdisplayshortskip{20pt}
	}
	\makeatother
	
	\usepackage{color}

	\usepackage{authblk}
	\title{\Large \bf Is Time Reversal in de Sitter Space a Spontaneously Broken Gauge Symmetry?}

	\author[1,2]{\Large Leonard Susskind}
	\affil[1]{LITP and Department of Physics, Stanford University, Stanford, CA 94305-4060, USA \vspace{1em}}
	\affil[2]{Google, Mountain View, CA, USA}
	\date{}
	\usepackage{upgreek}

\begin{document}
		
%\color{blue}	
		\begin{titlepage}
			\maketitle
			
			\begin{abstract}
			\Large

I'll begin  with some well-deserved acknowledgements:	
	I am grateful to Daniel Harlow for discussions of time-reversal holonomies. 
I have also benefited from a long ongoing correspondence with Edward Witten, but frankly in both cases I can't tell whether they agree with me or not.

I have often been accused of imprecision, especially toward the later parts of a paper, where I expect that my readers have ``caught on." That does eventually happen---the readers catching on and I thank them---but I'm now almost 86 and I can't wait.
So I've  tried to maintain a level of conceptual if not mathematical rigor throughout. 

Mathematical rigor(mortis) can sometimes be the enemy of conceptual clarity. I thank my friend Richard Feynman for  reminding me of that lesson.

Finally I thank the chatbot who gave me the definition of scaffold in section \ref{GS}. It was better than anything I was able to do.

\bn
		Symmetries of a  Holographic  theory; whether continuous or discrete, local or global,  are gauge symmetries of the bulk. This includes discrete space-time symmetries such as C and P.  But time-reversal is sufficiently different from other symmetries  that we may question the standard wisdom and ask whether symmetries involving T should be gauged in the bulk. Harlow and Numasawa \cite{Harlow:2023hjb} say yes; time-reversal is a gauge symmetry. Witten \cite{Witten:2025ayw}  says no: time reversal is different and does not manifest as a gauge symmetry of the bulk. My view is---yes---but with a  twist: Time-reversal is indeed a gauge symmetry; but it is	hidden by  spontaneous symmetry  breaking. In this paper I will review the case for spontaneous symmetry  breaking of time-reversal and explain the ``smoking gun"---a closed curve and a holonomy  which flips forward-going clocks to backward going clocks, and vice versa.

 \end{abstract}

		\end{titlepage}
		
		\rightline{}
		\bigskip
		\bigskip\bigskip\bigskip\bigskip
		\bigskip
		
		\starttext \baselineskip=17.63pt \setcounter{footnote}{0}

	\LARGE

		\tableofcontents
		
		\section{The Scaffold}\label{ Scaff}

\subsection{Static Patch Holography}\label{sph}

\begin{figure}[H]
		\begin{center}
			\includegraphics[scale=.2]{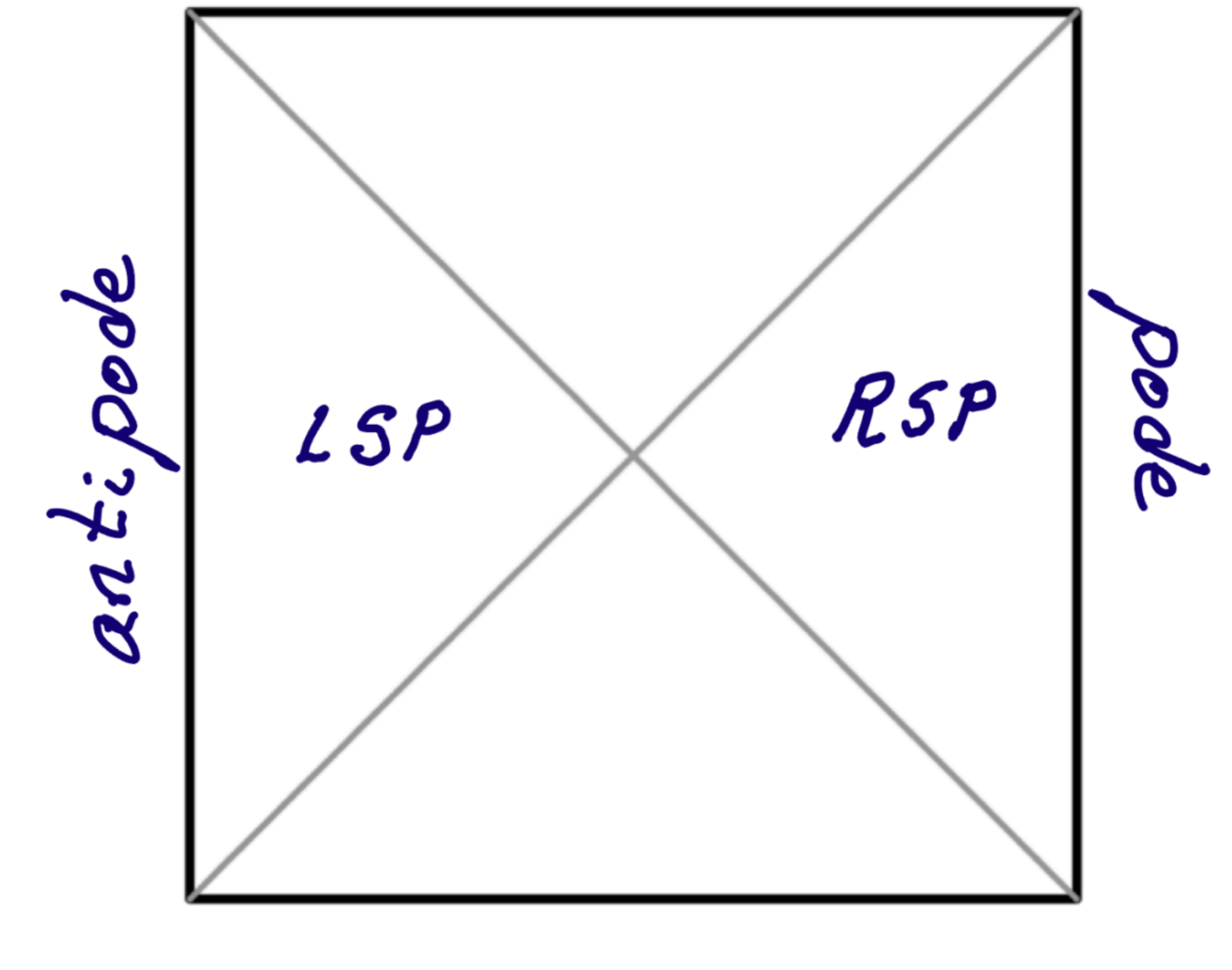}
			\caption{The de Sitter Penrose diagram showing the Right and Left Static patches together with the pode and antipode.}
			\label{square}
		\end{center}
	\end{figure}

I will follow the conventions and notations of \cite{Susskind:2026fjc}. 
The metric of de Sitter space in static coordinates is,
\be
ds^2 = -(1-r^2/\ell^2)dt^2 +\frac{dr^2}{(1-r^2/\ell^2)} +r^2 d\Omega_{(D-2)}^2.
\label{metric}
\ee
 The Penrose diagram is shown in figure \ref{square}. The horizon is at $r=\ell  $  and    the two points at $r=0$ are called the pode and the antipode. They lie at the centers of two opposing static patches---the pode at the center of the right  static patch and the antipode at the center of the left static patch (see figure \ref{square}  ).

The framework for this paper is static patch holography. We assume a Hilbert Space of dimension $2^N\times 2^N,$ a set of \dof called $\psi_L$ and $\psi_R,$  a Hamiltonian,
\be 
H= H_R -H_L
\label{H}
\ee
and a maximally entangled thermofield-double state. After tracing out the \dof \ of the left  static patch,  the right static patch is described by a Hilbert space of dimension $2^N,$
 degrees of freedom $\psi_r,$ a Hamiltonian $H_R,$ and a maximally-mixed infinite-temperature  density matrix. Expectation values are given by normalized traces,
 \bea  
 \langle\CO \ra \eq  \Tr \ \CO  \cr  \cr
 \Tr \eq \text{normalized trace}
 \label{trace}
 \eea

\subsection{Time-Reversal} \label{TR}

We will assume that  the holographic Hamiltonian  $H_R$ has an  antilinear  symmetry $\CT$ that includes both charge conjugation and time-reversal.  In the bulk the symmetry is identified with CRT---the product of charge conjugation, reflection of a single spatial axis, and time-reversal\footnote{Later, in section \ref{reveal} we will choose the reflection $R$ lie along an axis in the plane of the horizon.}. CRT is the generalization of CPT to  both odd and even
dimensional spacetime. When asking if time-reversal is a gauge symmetry we are  asking if CRT is gauged. 

Grouping charge conjugation with time-reversal is a convention due to Stueckelberg, Feynman, and Schwinger. The motivation is the idea that antiparticles are particles moving backward in time.

\subsection{Gauge Symmetry} \label{GS}

Symmetries of a holographic theory are realized as gauge symmetries of the bulk theory. Following \cite{Harlow:2023hjb} we will argue that this is true for $\CT.$

Gauge theories are scaffolds containing redundant elements which are useful in formulating the theory. 

\subsection*{Confession} 
Parts of the following definition comes from a chatbot. It is the only thing in this paper, or for that matter anything I have ever written, that does come from an AI.

\bn

\it  
Broadly defined, a scaffold is a temporary, supportive framework—physical, conceptual, or digital—designed to enable the completion of a task that would otherwise be beyond an individual's current capability. 
The scaffold is   eventually removed unless further work is to be done.
\rm
\bn

The redundant elements in a gauge theory include longitudinal and time-like gauge fields, gauge-variant matter fields including a possible Higgs multiplet, and most importantly, gauge symmetries. Gauge symmetries are not real symmetries but rather redundancies of the description. All physical states and observables are assumed to be invariant under a gauge  symmetry. If the gauge symmetry group is $G$ then all states $|\psi \ra$ and observables $\CO$ are constrained,
\bea 
G|\psi\ra \eq |\psi\ra  \cr \cr
G \CO G^{-1} \eq \CO
\label{const}
\eea
It is possible that a gauge theory can be written without the scaffold's redundancies;  purely in terms of gauge invariant quantities. There is a long history of such attempts but they are generally not useful: further work is always expected. When I speak of the holographic theory I always mean the scaffold theory.

Gauge symmetries can be continuous or discrete. U(1) and SU{N} are continuous gauge symmetries. The symmetry of the  Lenz-Ising model,  $Z_2$ can be gauged to give a discrete gauge symmetry.

\subsection{Gauge Symmetries: Local or Global} \label{LorG}

In the usual terminology the term global is used to mean a non-gauged symmetry, i.e., a symmetry not constrained by \eqref{const}. I don't think that is the right distinction to be made. Gauge symmetries---whether local or global---are always accompanied by gauge constraints, i.e.,   \eqref{const}. Un-gauged symmetries are not.

 I suggest using the distinctions: Locally and globally  gauged for all symmetries satisfying \eqref{const}; versus gauged and un-gauged for symmetries not satisfying \eqref{const}.

\bn

\it
 A global gauge symmetry is a global symmetry of the Hamiltonian---it acts simultaneously everywhere---with the added rule \eqref{const} requiring all states and observables to be invariant. 
 \rm
 
 \bn

In the Hamiltonian formulation of lattice QCD the gauge group is local; a product over all the lattice sites  of the local gauge group $SU(3).$
\be 
G_{QCD} =SU(3)^{\otimes N}  
\label{PiSuN}
\ee
where $N$ is the number of sites of the lattice.
The constraint equations \eqref{const} are the local nonabelian Gauss' laws. 

A global gauge invariance can arise from a local gauge theory by partial gauge fixing.  Whatever the reason for a global gauge invariance the gauge constraints \eqref{const} are global.

\subsection{A Tension} \label{tense}
Global gauge symmetries can be spontaneously broken. This  seems inconsistent with the  constraint \eqref{const} which demands that all states be invariant.
A state cannot be both invariant and not invariant.

 The resolution of this tension is to define spontaneous symmetry breaking not \ as \ a \ lack  \  of  \ invariance,  but as the presence \ of \ long \  range \ order. LRO  occurs when there is a breakdown of cluster decomposition, finite correlation persisting  to arbitrarily large distance. 
  The magnetized phase of an Ising ferromagnet is a good illustration. There are two states for each spin; $|u\ra$ and $|d\ra.$ The ground state $|GS\ra$ is the ``cat state,"
 \be  
 |GS\ra = |uuu\cdots u\rangle + |ddd\cdots d\rangle.
 \label{cat}
 \ee
 The cat state $|GS\ra$ is invariant under the $Z_2$ transformation
 \be
 Z_i  \to  -Z_i  
  \label{z2}
\ee
but it violates cluster decomposition,
\bea 
\langle Z_i \ra \eq 0  \cr   \cr
\langle Z_i Z_j \ra \eq 1,  \ \ \ \ \ |i -j| \to  \infty
\label{cldec}
\eea
Spontaneous symmetry breaking and  invariance can coexist but at the cost of  a breakdown of cluster decomposition. 

\subsection{LRO and Time-Reversal} \label{LRO}
Why  believe  in LRO for time-reversal? The argument begins  with a puzzle \cite{Susskind:2023rxm}. Consider a bulk   field operator $A$ far from the pode. For example the location of $A$ could be near the horizon, a distance $\sim \ell$ from the pode. In our world this distance would be of order $10^{10}$ light years but in some hypothetical universe it could be larger---arbitrarily large. From $A$ we construct 
an operator $C,$
\be 
C = [A,\frac{dA}{dt}]
\label{C}
\ee
(easy to remember: $C$ for commutator).

$C$ is odd under $\CT,$
\be  
\CT C \CT = - C
\label{Codd}
\ee
The expectation value of $C$ vanishes
\be 
\langle C \ra = \Tr C =0,
\label{C}
\ee
the trace of any commutator being zero. 

The puzzle is that when calculated  in the semiclassical limit---the limit of QFT in a fixed de Sitter background---the quantity $\langle C \ra$ is not zero\footnote{The operator $C$ is the derivative of $\langle [A(-t) , A(t)]\ra$ at $t=0$ which can be calculated by solving the wave equation for $A$. }. There is an apparent contradiction between the  trace-rule \eqref{trace} and the semiclassical bulk theory.

The resolution of the puzzle \cite{Susskind:2023rxm} was suggested in  \cite{Chandrasekaran:2022cip}. To obtain correlation functions that match the semiclassical limit requires the existence of a physical 
clock\footnote{The existence of the clock by itself is not enough. In addition the abstract time $t$ must be replaced by the clock-time $\tau$ as the argument of bulk degrees of freedom.}  located in the bulk (say at the pode)  to serve as a reference for ``dressing" operators. According to \cite{Chandrasekaran:2022cip} the presence of a small material clock at the pode changes $\langle C \ra$ from zero to its semiclassical value, even though  the distance between the clock and the location of $C$ may be arbitrarily large. This is the very definition of cluster breakdown.

\subsection{Forward-Going and Backward-Going Clocks}
Let's suppose that there is a material clock at the pode which registers  time $\tau. $
\footnote{In Von Neumann terms this has to do with the mathematical concept called the crossed product. 
As remarked in  \cite{Chandrasekaran:2022cip} the crossed product  is closely connected with  quantum frames of reference 
\cite{Aharonov:1967zza} }

It is natural to assume that $\tau$ is a monotonic function of $t,$ but is it monotonically increasing (a forward-going-clock FGC)?
$$\frac{d\tau}{dt} >0  \ \ \ \ \text{FGC}$$
Or    a  backward going clock (BGC)?
$$\frac{d\tau}{dt} <0  \ \ \ \ \text{BGC}$$
In a theory with time-reversal symmetry if a  FGC can exist  a BGC must also be possible.
Let us introduce projection operators,
\be 
\Pi, \ \Pi_f, \ \Pi_b
\label{PiPiPi} 
\ee
 constructed from bulk degrees of freedom located close to the pode. 
 $\Pi_f$ projects onto a state with a FGC; \   $\Pi_b$ projects onto a state with a BGC.

% Not only must the clock be present but one also must replace the abstract %time $t$ by the physical clock-time $\tau$ as the argument of the correlation %functions being studied. This means we must replace \eqref{C} by
%\be 
%C = [A,\frac{dA}{d\tau}]
%\label{Ctau}
%\ee
%Only when we do so will expectation values match their semiclassical c%ounterparts.

 The projection operators  satisfy\footnote{Note that  since $\CT$ includes charge conjugation a BGC is charge-conjugated relative to its FG counterpart.  },
\bea
\Pi &=& \Pi_f + \Pi_b \cr \cr
\Pi_b  &=&        \CT \Pi_f \CT   \cr  \cr
\Pi_f  &=&        \CT \Pi_b \CT
\label{Psat}
 \eea
(Remember $\CT^{-1} = \CT$)
 
 \bn
 
We also define the difference operator $\Pi_-$  
 \be
 \Pi_- = \Pi_f - \Pi_b 
\label{pi-} 
 \ee
 $\Pi_-$ is not a projection operator, it
is odd under $\CT$
 \be
 \CT  \ \Pi_- \CT = -\Pi_-
 \ee
 and  has eigenvalue $+1$ for states with a FGC and $-1$ for a BGC,
\bea 
\Pi_- \Pi_f \eq \Pi_f  \cr \cr
\Pi_- \Pi_b \eq - \Pi_b 
\label{prop}
\eea
 The operator $\Pi_-$ distinguishes FGCs from BGCs. 
 
\subsection{Dressing} 

The point of the operator $\Pi_-$ is to provide a reference for dressing observables.
  By multiplying $C$
  by $\Pi_-$ we turn $C$ from an operator odd under $\CT$ to a dressed operator even under $\CT.$   We will use a bar above an operator to indicate that it is dressed,
\be  
\bar{C} = C \Pi_-.
\label{barC}
\ee
  The dressed operator $\bar{C}$ is gauge invariant: Dressing operators is
 the same as making them gauge-invariant,
 $$\CT  \bar{C}   \CT  = \bar{C} $$
  or,
   \be 
 \CT C\Pi_- \CT =C \Pi_-
 \label{P-Ceven}
 \ee

According to  \cite{Chandrasekaran:2022cip} correlation functions of $\bar{C}$ match their semiclassical counterparts. They therefore satisfy cluster decomposition and in particular the expectation value of $\bar{C}$ is not zero.
\be 
\langle  \bar{C}  \ra \neq 0.
\label{barCneq0}
\ee

\subsection{Cluster Decomposition}

We also note  that,
\bea  
\langle C \ra \eq 0 \cr \cr
\langle \Pi_- \ra \eq 0
\label{cdno}
\eea
 The first of these follows from the fact that the trace of a commutator vanishes. The second from the assumed $\CT $ invariance of the holographic Hamiltonian---there are an equal number of FGC  and BGC states.
 Equations \eqref{barCneq0} and \eqref{cdno} together define a breakdown of cluster decomposition which  says, 
 \be
 \langle A B\ra \to  \langle A \ra \langle B \ra 
 \label{decomp}
 \ee
as the distance between $A$ and $B$ becomes arbitrarily large.
 
 Now let us return to the dressed operator $\bar{C},$
 According to \cite{Chandrasekaran:2022cip} the dressed versions of operators behave semiclassically.  Since semiclassically 
 $\langle C \rangle$ is non-zero it follows that,
  \be 
\Tr  C \Pi_- \neq 0.
 \label{dress}
 \ee
 
 On the other hand,
 \bea
 \Tr C \eq 0 \cr \cr
 \Tr \Pi_- \eq 0
 \label{Tr=0}
\eea
Since the locations of $C$ and $\Pi_-$ are arbitrarily distant from one another equation \eqref{Tr=0} constitutes a violation of cluster decomposition. Simply stated:

\bn
\it
  The presence of an arbitrarily distant clock has a significant effect on the expectation value of $C.$
  
  \rm
  \bn

Since both $\Pi_-$ and $C$ are $\CT$-odd it also constitutes a spontaneous breakdown of $\CT.$
 Witten expressed this\footnote{E. Witten, private communication} as follows:  ``The direction of the clock (at the pode) fixes the direction of time everywhere."

\section{Symmetry Revealed}\label{reveal}

In ordinary gauge theories like QED  combining  gauge theory with SSB leads to the Higgs phenomenon. Let's flesh that out a bit. Add to ordinary QED a charged scalar Higgs field $\phi$ whose magnitude is energetically constrained to be close to 1.
\be 
\phi \approx e^{i\eta}
\label{eieta}
\ee
Neither the electron field or the Higgs field is gauge invariant but the electron can be dressed to the Higgs field,
\be 
\bar{\psi} = \psi \phi^{*}
\label{dressel}
\ee
The composite field $\bar{\psi}$ is gauge invariant and     in the unitary gauge---the gauge in which $\phi$ is everywhere real---it is just $\psi$ itself.

\subsection{The Partially-Fixed Unitary Gauge}

 There is another gauge condition weaker than the unitary gauge defined 
  by partial gauge fixing. Instead of defining $\phi$ to be real everywhere we  define the gauge fixing condition to be that the phase $\eta$ is everywhere the same (but not necessarily zero). 
To give it a name I'll call it the partially-fixed unitary gauge. 
 In the partially-fixed unitary gauge there is a remaining global $U(1)$ gauge invariance. 
The overall phase of the Higgs field is a variable to be integrated over.

There is LRO and a breakdown of cluster decomposition in the partially-fixed unitary  gauge.
\be  
\langle \phi(x) \phi(y) \ra \to \text{finite}
\label{phiphi}
\ee
However whether or not  long-range order and SSB   take place  is a dynamical question; one that cannot be answered by simply defining the  partially-fixed unitary gauge.   There can be nonperturbative topological obstructions   to  the unitary (or partially-fixed unitary)  gauge.  A well-known example is   instantons  in the Polyakov $(2+1)$-dimensional   $U(1)$ model \cite{Polyakov}. Path integrating over a population of  instantons  can undo the LRO and lead to a confined phase in which correlations are short range. 

Let's just recall how that happens:
A single instanton does not lead to confinement or a decay of correlation functions. It takes a  gas of instantons, i.e., a sum over any number of instantons to  disorder the system.  This will come up later in section \ref{dlro} when we study holonomies.

\subsection{The Smoking Gun}\label{smoke}

Sidney Coleman once said that the Higgs effect leaves no track, by which he meant that there are no massless Nambu-Goldstone bosons or other degeneracies in the spectrum. Coleman argued that the Higgs effect is indistinguishable from no symmetry at all. If this were true  there would be no real meaning to $\CT$ being a gauge symmetry; no meaningful controversy over whether $\CT$ is a gauge symmetry or not.

But Coleman was not quite right\footnote{I am sure that Sydney knew this as well as anyone. Certainly as well as I did.}: The Higgs phenomenon leaves in its wake topological closed spacetime paths which, when traversed by parallel transport, reveal the symmetry operation as a holonomy.  For example when an electron is transported around an $SU(2)\times U(1)$ string it transforms into a superposition of electron and neutrino. One  may wonder if such a holonomy exists for $\CT;$ a holonomy that would exchange forward and backward going clocks. We will see that based on the principles of static patch holography---section \ref{sph}---the answer is yes.

\begin{figure}[H]
		\begin{center}
			\includegraphics[scale=.14]{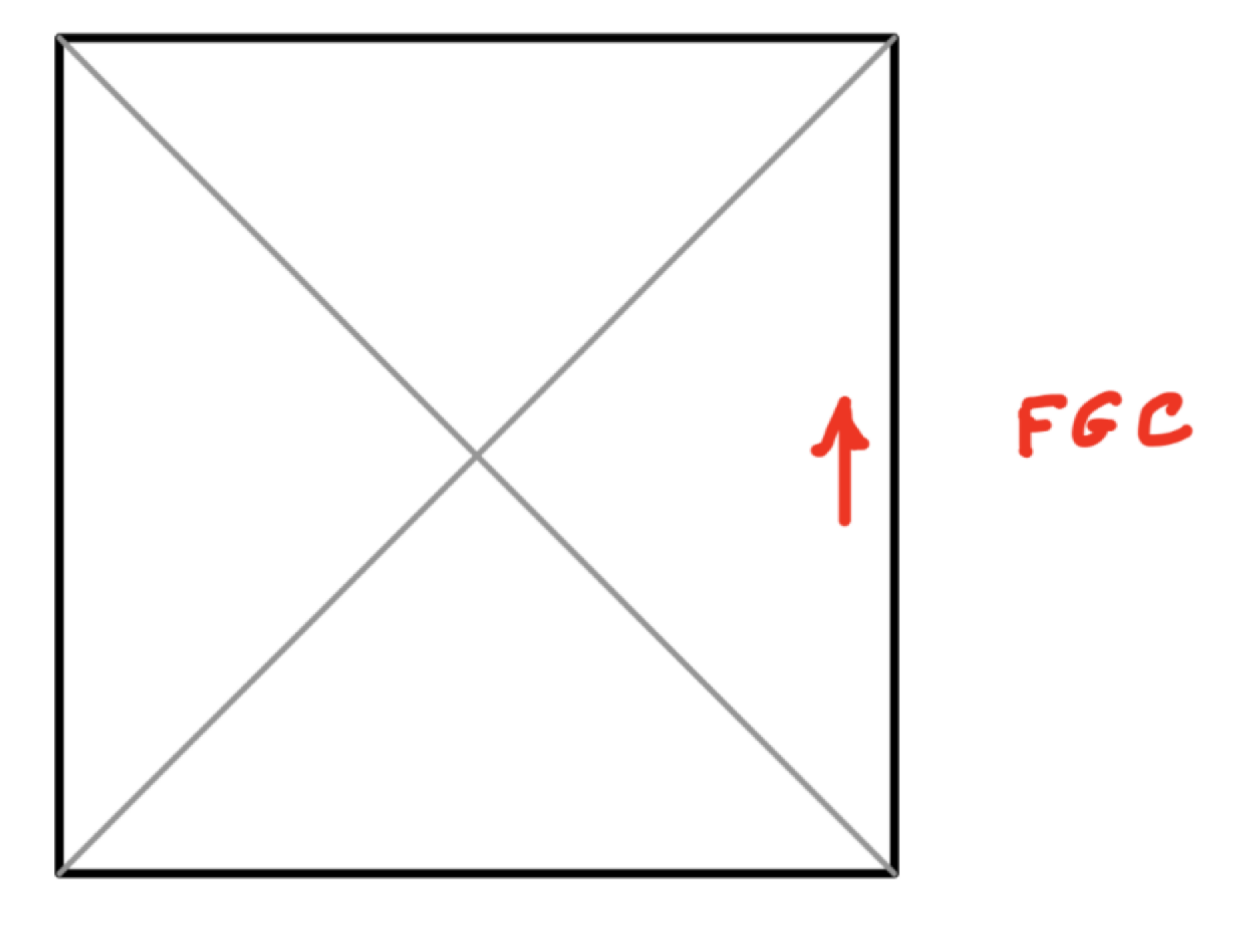}
			\caption{The de Sitter Penrose diagram and a Forward-Going-Clock located in the RSP.}
			\label{pen}
		\end{center}
	\end{figure}

For simplicity (and ease of drawing the diagrams) I will work with $(1+1)$-dimensional de Sitter space. However the arguments are easily generalized to higher dimensions (exercise for the reader). We begin with a Penrose diagram showing two entangled static patches together with a forward-going clock (FGC) at the pode (figure \ref{pen}).

\bn
Each point of a Penrose diagram represents a $(D-2)$-sphere,  in the case $D=2$ a zero-sphere. A zero-sphere  means a pair of points; the solution to the equation $x^2 = a^2.$

\begin{figure}[H]
		\begin{center}
			\includegraphics[scale=.22]{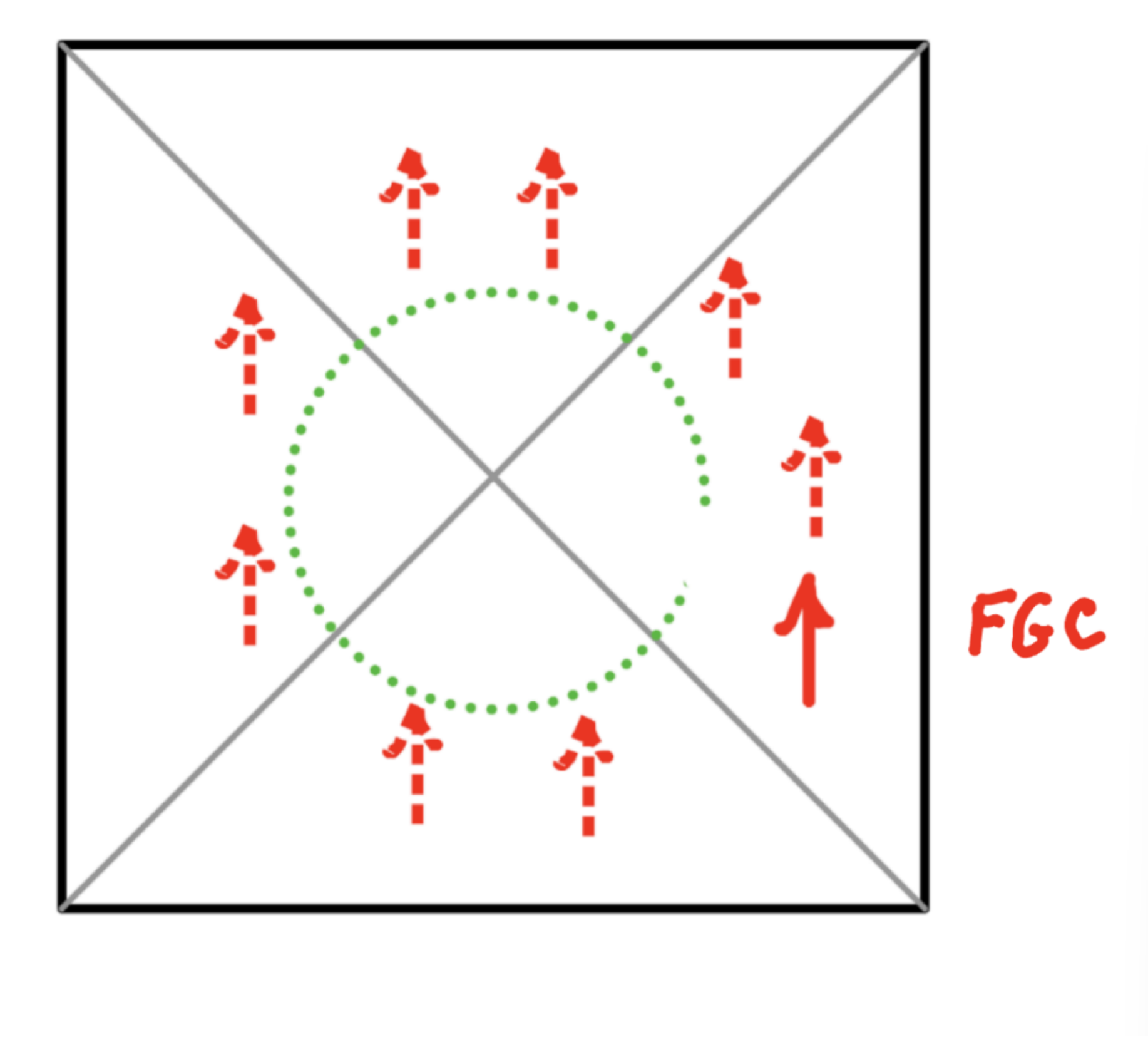}
			\caption{Transporting the FGC around the bifurcate horizon. But something important has been left out.}
			\label{loop}
		\end{center}
	\end{figure}

\begin{figure}[H]
		\begin{center}
			\includegraphics[scale=.23]{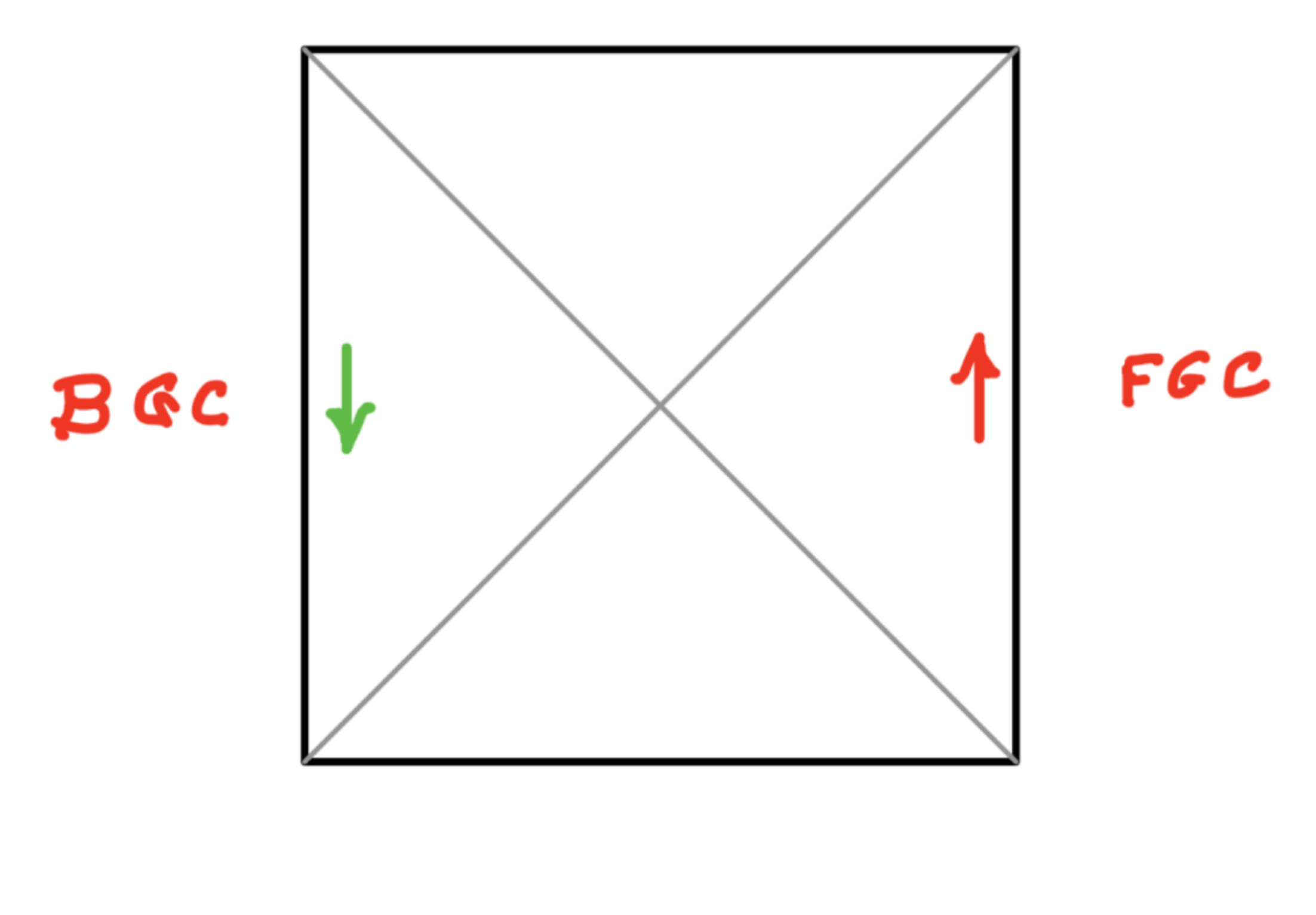}
			\caption{}
			\label{entang}
		\end{center}
	\end{figure}

\begin{figure}[H]
		\begin{center}
			\includegraphics[scale=.23]{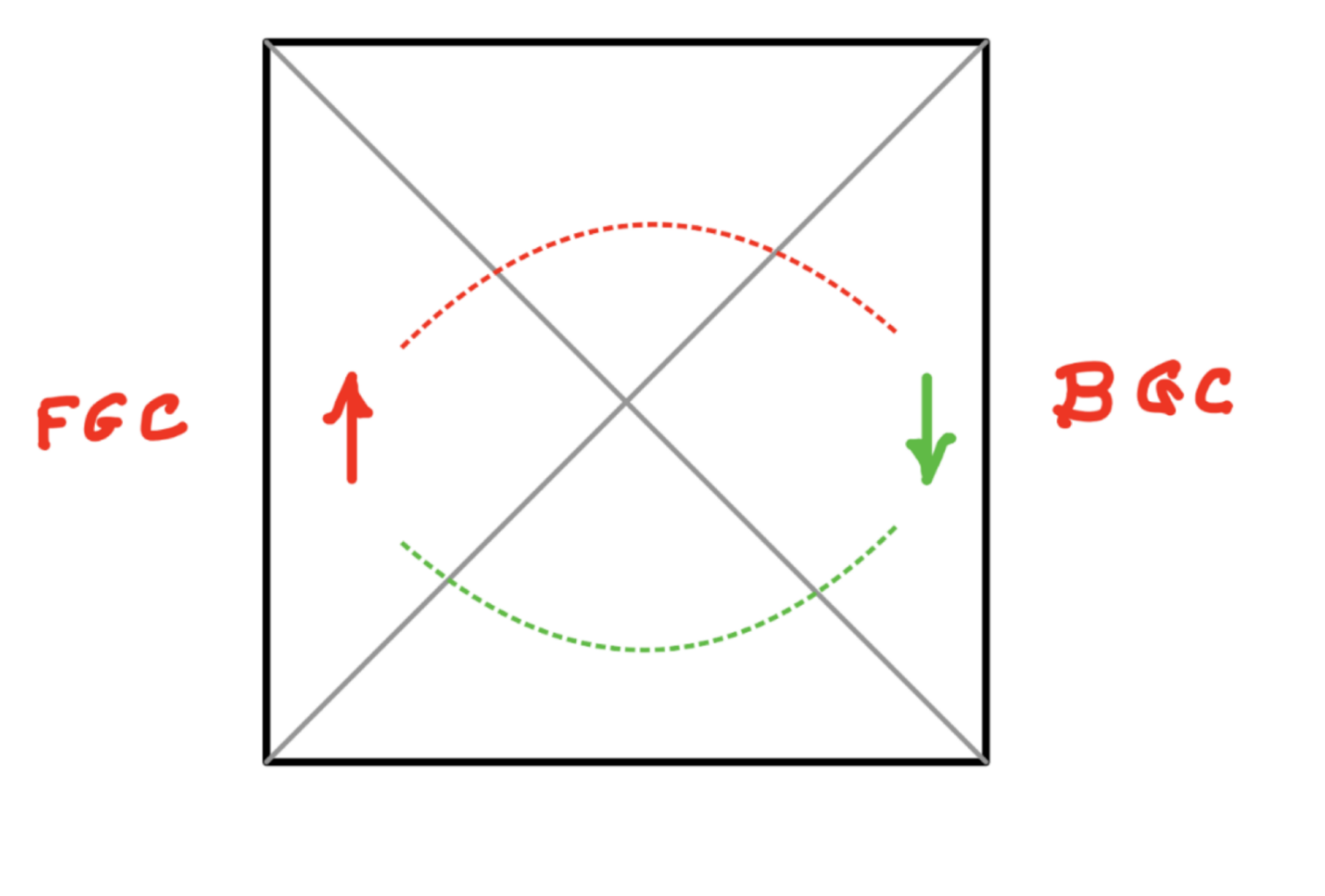}
			\caption{Maximum entanglement implies that if a FGC is present  at the pode a BGC is automatically present at the antipode.}
			\label{cycle}
		\end{center}
	\end{figure}

Now let's parallel-transport the FGC around the bifurcate horizon bringing it back to its original spacetime location as in figure \ref{loop}. 
	\bn
				It seems that the FGC returns to a FGC, i.e., no nontrivial holonomy.	
But something important has been left out; the left and right static patches are maximally entangled. In projecting onto a state with a FGC at the pode we automatically project onto a BGC at the antipode (figure \ref{entang}). 	
	When the FGC is transported to the antipode the BGC will be transported to the pode along a trajectory that mirrors the trajectory of the FGC as shown in figure \ref{cycle}. 
\bn
The FGC at the pode has transformed into a BGC.	

		Geometrically this may be described as follows:

	\bn
	\it 
		Maximal entanglement diametrically identifies the geometry across the bifurcate horizon. In the Euclidean
theory the disc is replaced by a cross-cap. This
means that the circuit shown in figure 4 is a complete circuit—
not just half a circuit. The FGC at the pode has been exchanged for a BGC
thereby revealing the time-reversal symmetry hidden by
Higgs.
\rm

\bn
The actual (Euclidean) geometry is not just a cross-cap. It is a ``framed"  cross cap in which each point has an attached framing vector. The framing vector  represents the two points comprising  the zero-sphere. The holonomy represented by figure \ref{cycle} exchanges the two points thus changing  the
orientation of the framing vector.  In other words  at each point of the Penrose diagram it exchanges the two points representing the zero-sphere.
The reflection of the framing vector is the $(1+1)$-dimensional version of a reflection in the plane of the horizon\footnote{The framing direction can be thought as the limit of a very small additional oriented direction. In the $(1+1)$-dimensional case   the 2-dimensional framed cross-cap is the limit of a three-dimensional geometry. The cross-cap is an unorientable geometry  but the framed cross-cap is orientable. This is fortunate for fermions which would otherwise have difficulties.}. 

\begin{figure}[H]
		\begin{center}
			\includegraphics[scale=.23]{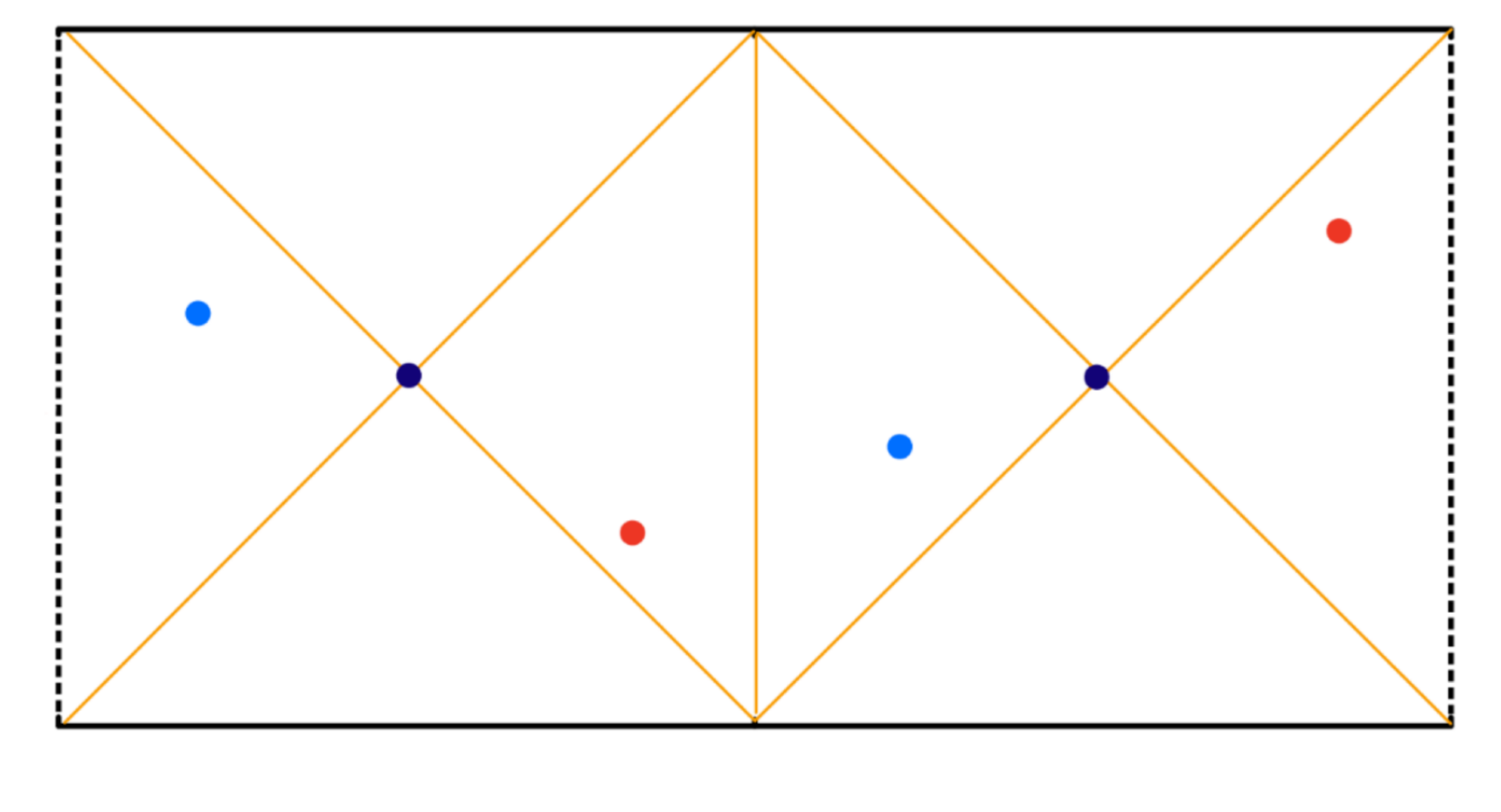}
			\caption{The full conformal diagram for $(1+1)$-dimensional de Sitter space. Dots of the same color are identified as the endpoints of framing vectors. }
			\label{rect}
		\end{center}
	\end{figure}

Does all of this mean that the bifurcate horizon is a
special or singular place? The answer is no. It is
easily  seen in the example of (1+1) dimensional de Sitter
space where every point in the Penrose diagram represents
a pair of points. Each pair is the  horizon for some 
observer. This can be made clear by doubling the Penrose diagram and imposing periodic boundary conditions. 
The  full conformal diagram for $(1+1)$-dimensional de Sitter space is shown  in figure \ref{rect}.

	The points in figure \ref{rect} come in pairs which in the Penrose diagram (figure \ref{pen})
are identified. The pairs are
identified by matching colors; the black pair; the red
pair; the blue pair and so on.
Every matched pair is the horizon
of a unique static patch. There is nothing special about
any given pair just as there is nothing special about a
given static patch.

\begin{figure}[H]
		\begin{center}
			\includegraphics[scale=.24]{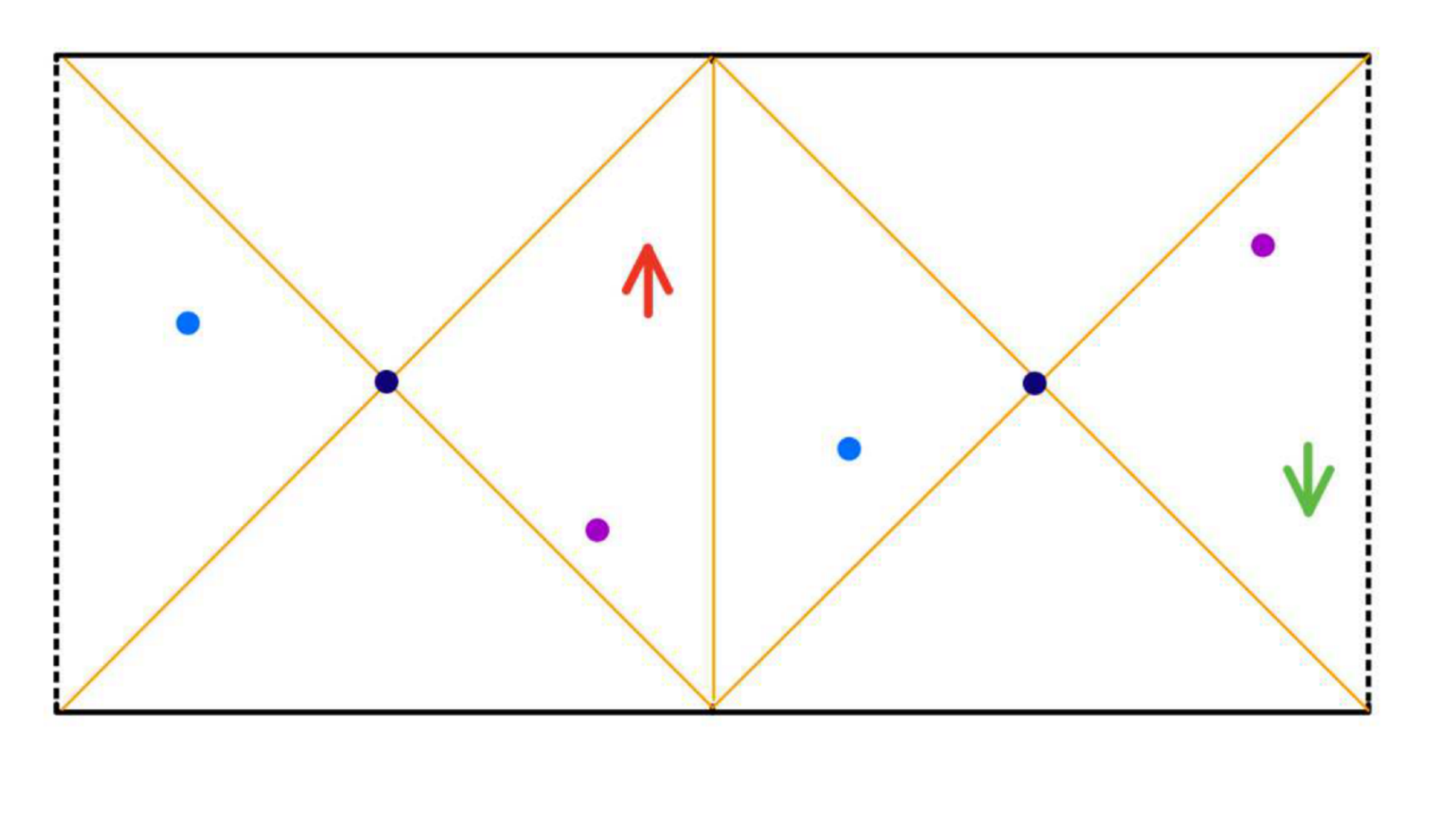}
			\caption{Adding a FGC and a BGC to the conformal diagram.}
			\label{redgreen}
		\end{center}
	\end{figure}

In figure \ref{redgreen} we add to the conformal diagram a BGC in
green and its FGC image—required by maximal entanglement—
in red.

\begin{figure}[H]
		\begin{center}
			\includegraphics[scale=.23]{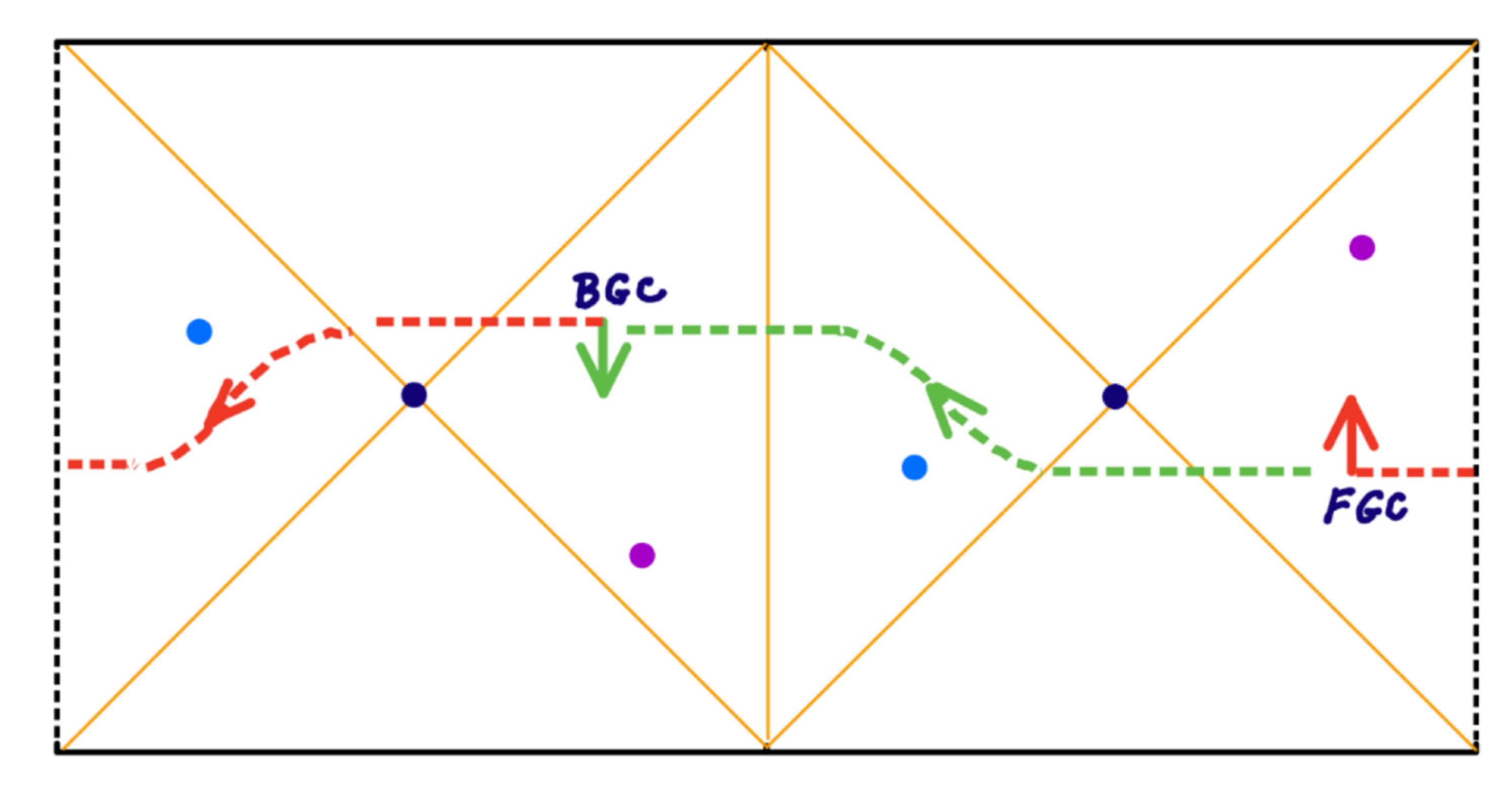}
			\caption{A closed loop and a holonomy  that exchanges the FGC and BGC.}
			\label{fin}
		\end{center}
	\end{figure}

 Now form a closed loop in the identified space by
parallel-transporting  the green BGC to the location of
the FGC while at the same time transporting the red
FGC to the position of the original BGC. The contour
of this second transport is the image of the first. It
brings the red FGC to the location of the original BGC
as shown in figure \ref{fin}.

In short, parallel transport along the closed path ex-
changes the original BGC for a FGC. The holonomy is
CRT, i.e.,   $ \CT $. The hidden time-reversal symmetry is
thus revealed.

In this case there is no localized vortex-like defect.  The closed
contour  surrounds the bifurcate horizon; for example focus on the horizon defined by the black points in figure \ref{fin}. The full contour (consisting  of the green and red dashed portions goes below the black dot on the right side of the diagram, and above the black dot on the left side. When drawn as a Penrose diagram figure \ref{fin} becomes figure \ref{cycle}.

However, from figure \ref{fin} it is clear that the same is  
true for every static patch, i.e., for every pair: there is nothing special about the black pair. 
The localized
defect is replaced by the topology of the global geometry.

\subsection{Does the Holonomy Cause Disorder?}\label{dlro}
LRO can be undone by summing over 
  instantons, starting with zero and going to arbitrarily large number.   A single instanton does not have such an effect. The holonomy that we exhibited in \ref{smoke} is the analog of a single somewhat delocalized  instanton,  centered at the bifurcate horizon. It is a fixed  feature of the de Sitter  geometry and the maximal entanglement of the thermofield-double state, not to be summed over.  The mechanism for disordering (confining) the theory is not present.

\section{ Is the Gauge Invariance Local? }

I argued earlier that $\CT$ is a global, not a local gauge invariance. However this is not entirely
true. The full two-sided system including the left and
right static patches in figure \ref{pen}  is described by a product
Hilbert space and a Hamiltonian
$H_R - H_L$
where R and L refer to the left and right opposing static
patches. $ H_R$  and $H_L$ commute with each other and are
composed of separate degrees of freedom. Each side has
its own time-reversal symmetry, $\CT_R$ and $\CT_L.$
\be 
\CT = \CT_R  \times   \CT_L.
\label{TeqTT}
\ee
$\CT_R $ and $\CT_L$ act locally on the left and right holographic degrees of freedom. In that sense there is an element of locality to the $\CT$ gauge invariance.

 However these individual local symmetries of the Hamiltonian---$\CT_L$ and $\CT_R$---
are spontaneously broken by the maximally entangled TFD 
state of the two-sided system.

\section{Conclusion}

I've argued that   CRT in de Sitter space is indeed a
gauge symmetry as forseen by Harlow and Numasawa,
but with a twist; it is hidden by spontaneous breaking.
The only direct evidence for the symmetry is the exis-
tence of loops that surround the bifurcate horizon and
give rise to a holonomy exchanging  FGCs and BGCs.

Gauge-variant operators like $C$ and $\Pi_-$ 
do not satisfy cluster decomposition but their dressed, or gauge 
invariant versions do. The correlation functions of
dressed operators are expected to agree with semiclassical
expectations in the limit
$$\frac{\ell}{\ell_{planck}}  \to \infty. $$
This was expressed in \cite{Chandrasekaran:2022cip} by saying that only dressed (gauge invariant) operators are observable and have limits that agree with semiclassical expectations.
Among other things this means that dressed operators satisfy cluster decomposition (no LRO). 

If it were not for the holonomy exhibited in section \ref{smoke} there would be no ``track" left by the spontaneous symmetry breaking; it would be as if there was no symmetry at all, which I think is what Coleman and Witten had in mind.

	\end{document}